\documentclass[epj,nopacs]{svjour}
\usepackage{graphicx,epsfig,mathrsfs,amsmath}
\def\rmd{\mathrm d}
\def\rmi{\mathrm i}
\def\rme{\mathrm e}
\def\be#1\ee{\begin{equation}#1\end{equation}}
\newcommand{\ba}{\begin{eqnarray} }
\newcommand{\ea}{\end{eqnarray} }
\begin{document}
\title{Relativistic invariance of the vacuum}
\author{Adam Bednorz\inst{1}
}                     
\offprints{}          
\institute{Faculty of Physics, University of Warsaw,
ul. Ho\.za 69, PL00-681 Warsaw, Poland}
\date{Received: date / Revised version: date}
%
\abstract{
Relativistic invariance of the vacuum is (or follows from) one of the Wightman axioms which is commonly believed to be true. Without these axioms, here we present a direct and general proof of continuous relativistic invariance of all real-time vacuum correlations of fields, not only scattering (forward in time), based on closed time path formalism. The only assumptions are basic principles of relativistic quantum field theories: the relativistic invariance of the Lagrangian, of the form including known interactions (electromagnetic, weak and strong), and standard rules of quantization. The proof is in principle perturbative leaving a possibility of spontaneous violation of invariance. Time symmetry is however manifestly violated.
} 
\maketitle
\section{Introduction}

Relativistic invariance of the vacuum is a basic property of all quantum field theories based on relativistically invariant dynamics, directly following from one of Wightman axioms \cite{cpt}.  The (third) Wightman postulate assumes the existence of a single ground state with zero energy-momentum eigenvalue \cite{jost}. 
The postulate is  generally accepted by high energy community \cite{cole} although the problem is more sophisticated than it looks and the validity of the axiom itself has been questioned \cite{segal,koks}.
However, to postulate zero energy is in conflict with renormalization which may add indefinite contribution. Moreover, this postulate should be redundant since the vacuum is already determined by the Lagrangian and the rules of quantization.
Indeed, without the Wightman axiom, the invariance has been already proved for special cases of relativistic quantum field theories: single fields (electromagnetic, current, etc.), free theories, Feynman (forward) time-ordered (in-out) correlations of fields, scattering processes, second order and some other classes of correlations\cite{lors,qft,weinb,peskin,bell-book}, but never in general (for arbitrarily time-ordered correlations). A large part of high energy community wrongly claim it is already proved in general \cite{nair}. For instance, a naive proof by reconstruction of vacuum wavefunction out of Feynman correlations fails because such a reconstruction is incomplete and not unique.
A general proof requires to show Lorentz invariance of all arbitrarily ordered correlations at real spacetime points.
The invariance could be certainly proved only if the underlying dynamics is invariant
and conserved charges are zero. It has been suggested that Nature
 may violate the invariance directly (by non-invariant dynamics) or spontaneously (by a Higgs-like mechanism) \cite{kost}. We do not discuss these two last possibilities here.

In this paper, we confirm the common intuition by a direct proof, not relying on Wightman axioms. Although formally general, 
it is better to understand it as perturbative. We show that zero-temperature vacuum correlations of fields at real spacetime points are invariant under continuous transformations of a reference frame. 
We shall use the framework of the closed time path formalism (CTP) \cite{schwinger,kaba,kel} where correlations are defined on the complex path (going downwards with respect to imaginary part). It is defined in a particular reference frame, hence the invariance is not manifest.
We only assume basic, accepted principles of relativistic quantum field theories: (i) relativistically invariant Lagrangian, including known interactions (electromagnetic, weak and strong, specified in detail later in the paper) and
(ii) rules of quantization (standard construction of field expectation values, consistent with CTP).
However, the discrete transformations like time-reversal and parity have to be treated separately. If the underlying dynamics is invariant under charge
conjugation or parity reflection then the proof remains valid for these transformations \cite{cpt}. Unfortunately, even if the dynamics is symmetric
with respect to time reversal, then CTP is not (except special cases, e.g. at space-like points or in scattering problems).
In the same way, CTP is in general not invariant under joint charge-parity-time reversal in contrast to dynamics.
It it closely related to violation of time symmetry in quantum noninvasive measurements \cite{our}.

The paper is organized as follows. We first recall the construction of the complex time path correlations, defining relevant correlations of fields. Next, we show that the correlations are independent of the particular shape of the path. Then we prove the Lorentz invariance of the zero-temperature no-charge vacuum for continuous transformations.
Then we discuss possible pitfalls in other attempts of the proof based e.g. on momentum representation, which is dedicated to all who claim that the invariance is trivial or easy extension of existing proofs \cite{nair}. Finally, we will see
that  time symmetry is absent in CTP.
Throughout the paper, we use the convention $\hbar=c=1$ and $\beta=1/k_BT$ (inverse temperature).

\section{Closed time path formalism and definitions}
\label{ctps}

\begin{figure}
\includegraphics[scale=.45]{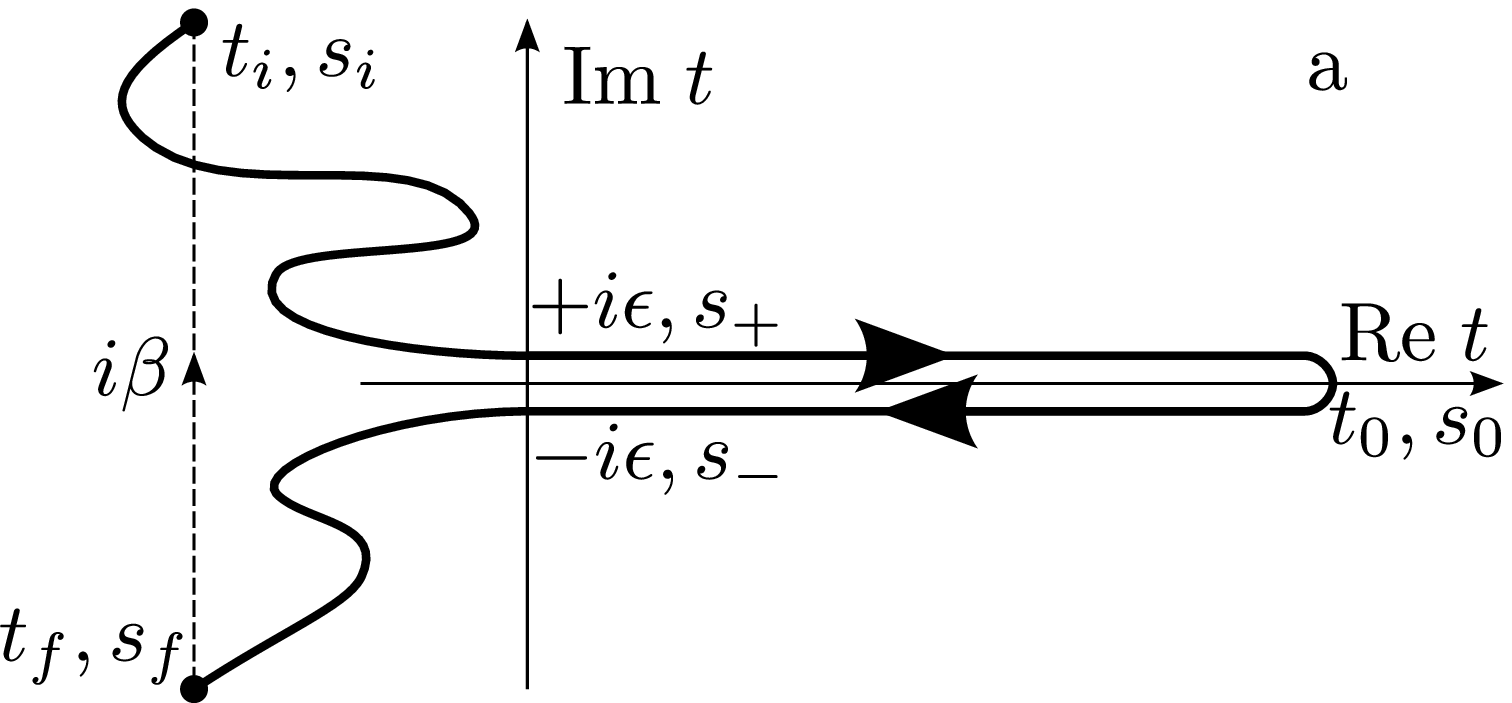}
\includegraphics[scale=.45]{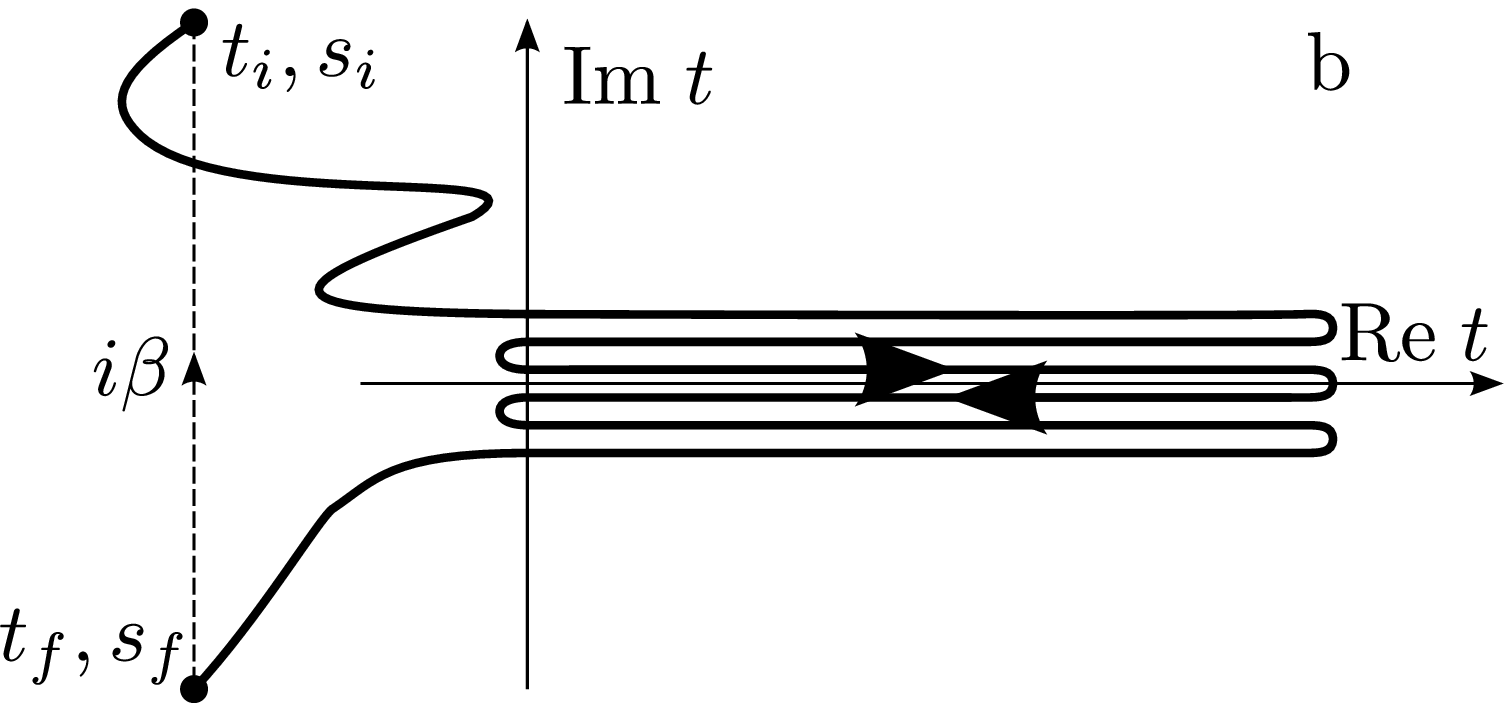}
\caption{The closed time path in the complex time plane. 
All the path cannot go upwards in the imaginary direction. The finite temperature implies the jump $\rmi\beta$, which extends to infinity
at zero temperature. Apart from the above conditions, the shape of the path is arbitrary. We stress that the path is ordered by
the real parameter $s$. All the important times have been marked. The shift between real parts is infinitesimal only
for better visibility but in fact can be zero.
(a) The simple in-out-in (shortly in-in) contour with a single pair of real parts (between $0$ and $t_0$).
(b) The generalized contour for arbitrarily ordered correlations.
}\label{ctppcon}
\end{figure}

The complex closed time path $C$ is a specially chosen curve of time $t$ in the complex plane as a function of a real parameter $s$.
The curve is continuous with nonincreasing $\mathrm{Im}\,t$ (important for convergence of integrals, involving energies bounded from below and open from above) except a jump of $\rmi\beta$, which becomes imaginary infinity at zero temperature, see Fig. \ref{ctppcon}. The physical part of the contour is located on the real axis (both upper and lower).
In particular, the curve in Fig. \ref{ctppcon}a is
parametrized by real $s\in[s_i,s_f]\to t(s)$ with 
\ba
&&\mathrm{Im}(\rmd t/\rmd s)\leq 0,\label{ctpd}\: t(s_f)=t(s_i)-\rmi\beta,\\
&& \mathrm{Im}\, t\to 0_+\mbox{ and } \mathrm{Re}\,\rmd t/\rmd s>0\mbox{ for }s\in [s_+,s_0],\nonumber\\
&& \mathrm{Im}\, t\to 0_-\mbox{ and } \mathrm{Re}\,\rmd t/\rmd s<0\mbox{ for }s\in[s_0,s_-],\nonumber\\
&& t(s_\pm)=0_\pm, t(s_0)=t_0,\:s_i<s_+<s_0<s_-<s_f.\nonumber
\ea
The generalized contour in Fig. \ref{ctppcon}b contains multiple real parts.

To facilitate an easier notation throughout the paper we define Heaviside and Dirac functions for
complex time arguments $t$, 
\ba
&&\theta(t-t')=\theta(s-s'),\:\delta(t-t')=\delta(s-s')/(\rmd t/\rmd s),\nonumber\\
&&\partial_t=(\rmd t/\rmd s)^{-1}\partial_s,\:\rmd t=(\rmd t/\rmd s)\rmd s
\label{tcom}
\ea
 where as usual $\theta(s-s')=1$ for $s>s'$ and $0$ otherwise
while $\int \rmd s\delta (s-s')=1$ and $\delta(s-s')=0$ for $s\neq s'$. Therefore $\theta$ defines a special order of times
along the path, and it may be converse to the usual order.
This is important since only $s$ is the real, ordering parameter, while $t$
 is complex and thus has no natural ordering.
Note that in all contour time-integrals we essentially integrate over $s$.

Let us take the generating functional for quantum correlations (often referred to as Green functions) with respect to the action, making use of the Lagrangian approach, namely \cite{kaba,bell-book,feyn,mats,kap,bam,mills,calz,wein,kobeskow,kra,mabi}
\ba
&&G\equiv\left\langle \prod_k A_k(t_k)\prod_l\phi_l(t_l)\right\rangle=\nonumber\\
&&
\left.\prod_k\frac{\delta}{\delta \rmi\chi_k(t_k)}\prod_l\frac{\delta}{\delta \rmi\eta_l(t_l)}\rme^{S[\chi,\eta]}\right|_{\chi=0,\eta=0}\label{gree}
\ea
where
\be
\rme^{S[\chi,\eta]}=\frac{\int DAD\phi\,\rme^{\int_C \rmi \rmd t(L+\sum_k(\chi_k(t)A_k(t)+\eta_k(t)\phi_k(t))}}
{\int DAD\phi\,\rme^{\int_C \rmi\rmd t\,L}}\label{genn}
\ee
with auxiliary commuting ($c$-) field $\chi$ ($\chi\chi'=\chi'\chi$) and anticommuting Grassmann field $\eta$ ($\eta\eta'=-\eta'\eta$).  Here $t_k$ can be located everywhere
on the path (upper real, lower real, left complex part).
In the above, $A$ and $\phi$ are  generic bosonic and fermionic (Grassmann) fields, respectively,
and the Lagrangian is a time-local function of fields and their time derivatives, namely
$L=L[A,\dot{A},\phi,\dot{\phi}]$. The integral $\int_C \rmd t$ is over the just defined time path,
and should be read as an integral over real $s$, $\int \rmd s (\rmd t/\rmd s)$. The boundary conditions for fields are
$A(t_i)=A(t_f)$ and $\phi(t_i)=-\phi(t_f)$, for bosonic and fermionic field, respectively (Kubo-Martin-Schwinger condition \cite{kms}).
Standard (in-out) correlations are obtained for the auxiliary fields nonzero only on the upper real part of the contour.
General correlations involve both parts of the contour (in-out-in or in-in) or even more (in-in-in, etc.).
The path-integral formulation of CTP is equivalent to Hamiltonian formulation, as we show in Appendix \ref{appa}.
The standard issues such as Wick theorem, conservation laws and renormalization are resolved in CTP analogously to
in-out correlations, see Appendix \ref{techa}.

\section{Contour shape independence}
\label{shape}

The fact that we can freely wiggle the contour is reasonably believed for CTP \cite{mills,ksw,nieg,evans}, but
a direct proof is very instructive \cite{gelis}. 
Let us prove that (\ref{genn}) is independent of a particular shape of CTP, because the proof of Lorentz
invariance will be analogous.
It follows from the invariance of $DA$ and $D\phi$ under infinitesimal transformation $\delta t(s)$
parametrized by real $s$ (the variation may change endpoints but the jump must be kept constant)
\ba
&&A(t)\to \rme^{\delta t\partial_t}A(t)=A(t)+\delta t\partial_t A(t)+\dots,\nonumber\\
&&\phi(t)\to \rme^{\delta t\partial_t} \phi(t)=\phi(t)+\delta t\partial_t \phi(t)+\dots
\ea
which gives $L(t+\delta t)=\exp(\delta t\partial_t)L(t)$. This means that the effect of the contour changes
can be absorbed into a shift under time integral and a such it disappears (analogously to an integral over a total derivative). We stress that this proof is valid only if do not change the shape of
the real part of the contour, except shifting $t_0$ if it lies beyond all relevant times. It is essential that the fields
are smooth.

An alternative, more explicit proof relies on perturbative approach. One starts with Gaussian, diagonalized Lagrangian
\ba
&&L_0 = \sum_k \left[\psi^\ast_k(i\partial_{t}-E_k)\psi_k
      +  A_k^\ast(i\partial_{t}-E_k)A_k\right.\nonumber\\
&&\left.+\dot{B}_k^2/2-E_k^2B_k^2/2)\right].
\ea
In the above, we have denoted complex $\psi$ instead of real $\phi$ and distinguished the
linear time-derivative bosonic field $A$ from the quadratic $B$.
The two-point Green function  for $B$ reads
\be
\langle B_k(t)B_l(t')\rangle_0=\frac{\delta_{kl}}{2E_k}\sum_\pm 
\mp \frac{\rme^{\pm i|t'-t|E_k}}{1-\rme^{\pm \beta E_k}},\label{bbgreen}
\ee
where $|t'-t|=(t-t')\theta(t-t')+(t'-t)\theta(t'-t)$.
For  $A$ we get (\ref{bgreen}) and for $\psi$  -- (\ref{fgreen}).

If we add interaction term, $L=L_0+L_I$, and $L_I$ depends on fields appearing in $L_0$, the generating function (\ref{genn}) -- by virtue of the Wick theorem (\ref{wicka}) -- will be a sum of integrals represented by
Feynman diagrams, with lines (propagators) corresponding to free Green functions (\ref{bgreen}), (\ref{fgreen}) and (\ref{bbgreen})
and vertices corresponding to $L_I$.
Suppose that $t(s)$ is varied by infinitesimal $\delta t$. Let us consider the variation of Green functions.
Note that
\be
\delta\langle X(t)X^\ast(t')\rangle_0=\left(\delta t\partial_t+\delta t'
\partial_{t'}\right)\langle X(t)X^\ast(t')\rangle_0\label{dfree}
\ee
where $X=A,B,\psi$ and subscript $0$ denotes averaging (\ref{gree}) with respect to $L_0$.
The free Green function depends only on endpoint times $t$,$t'$, not the shape of CTP between them. 

Now, each physical diagram corresponds to average of the type
\be
\int_C \rmd t_1\cdots \rmd t_N\left\langle\prod_{n\leq N} L_I(t_n)\prod_{k>N} X_k(t_k)\right\rangle_0,\label{tt0}
\ee
where $X=\psi,A,B$ and $t_k$ lie in the real (right hand) part of the contour for $k>N$. So we get
\ba
&&\delta\int_C \rmd t_1\cdots \rmd t_N\left\langle\prod_n L_I(t_n)\prod_k X_k(t_k)\right\rangle_0=\\
&&
\int \rmd s_1\cdot\cdot\rmd s_N\sum_{m\leq N}\frac{\partial}{\partial s_m}\left\langle\delta t_n \prod_{n\leq N} L_I(t_n)\prod_{k>N} X_k(t_k)\right\rangle_0\nonumber,
\ea
where we used Wick-decoupling into products of free Green functions, Leibniz rule $\delta(AB)=B\delta A+A\delta B$, 
$\delta \rmd t=\rmd\delta t=\rmd s\partial_s\delta t$,  (\ref{dfree}), inverse Leibniz rule applied to $\partial_s$ and
back Wick-coupling.
The final result of the shape variation integral is zero, which completes the proof.

\section{Lorentz invariance}

It is natural to expect that Green functions are invariant under relativistic Lorentz transformations at zero temperature
if the Lagrangian is Lorentz-invariant.
However, starting from CTP makes it not obvious because the contour prefers some time direction. Intuitively, we expect that this
should not bring about any problem but the warning light comes already from the non-invariance due to finite temperatures, which
enters only as a jump in the contour. Below, we show that the Green functions at real times in zero-temperature vacuum are indeed invariant under
continuous  Lorentz transformations in Lorentz-invariant field theories but not necessarily under time reversal.

\subsection{Relativistic notation in CTP}

Let us briefly recall relativistic notation.
The components of position four-vector are denoted $x^\mu$, $\mu=0,1,2,3$,
with time $x^0=t$ and three spatial coordinates $x^k$, $k=1,2,3$. Fields, functions of position will be denoted as $f(x)$ where $x=(x^0,x^1,x^2,x^3)$. 
The derivatives are denoted by $\partial_\mu=\partial/\partial x^\mu$. From now on, Greek indices are reserved for four-vectors
and higher Lorentz-like tensors and structures.
It is extremely important for the purpose of this paper, that the time $x^0(s)$ is here a complex function of the real parameter $s$.
In particular the time derivative $\partial_0$ is translated into the real derivative $\partial_s$
by
\begin{equation}
 \partial_0=(\rmd x^0/\rmd s)^{-1}\partial/\partial s
\label{timeim}
\end{equation}
Similarly, we introduce Dirac four-delta, including complex time,
\be
\delta^4(x-x')=\frac{\delta(s-s')}{\rmd x^0/\rmd s}
\delta(x^1-x^{\prime 1})\delta(x^2-x^{\prime 2})\delta(x^3-x^{\prime 3}).
\ee
and four-integral $\int_C \rmd^4x$ with $\rmd^4x=\rmd x^0\rmd x^1\rmd x^2\rmd x^3$ and $\rmd x^0=(\rmd t/\rmd s)\rmd s$. 
For a compact notation of relative four position including the complex-time component of the four-vector 
on CTP we will denote $|x-x'|^\mu$ with $|x-x'|^{1,2,3}=(x-x')^{1,2,3}$ but 
\be
|x-x'|^0=\theta(x^0-x^{\prime 0})(x^0-x^{\prime 0})+\theta(x^{\prime 0}-x^{0})(x^{\prime 0}-x^{0}),
\label{x00}
\ee
where $\theta(x^0-x^{\prime 0})=\theta(s-s')$ applies to complex $x^0$ and $x^{\prime 0}$ along CTP.

The rest of conventions, including Lorentz generators $J$, spinors and Lagrangian density $\mathcal L$ (becomes Lagrangian $L$ when space-integrated) are consistent with textbooks \cite{qft,peskin,itzyk},
which we recall in Appendix\ref{lorb} for completeness.

The essential \emph{postulate} of QFT is that the Lagrangian density is a Lorentz-invariant.
It means that 
the generating functional (\ref{genn}) can be written now as
\ba
&&\rme^{S[\chi,\xi,\eta]}\propto\label{genl}\int DADBD\bar{\psi}D\psi\;\times\\
&&\exp\int_C \rmi\rmd^4x\left(\mathcal L+\sum_k(\chi_kA_k+\xi_k\cdot B_k+\bar{\eta}_k\psi_k+\bar{\psi}_k\eta_k)\right)
\nonumber
\ea
where all fields are functions of spacetime, e.g. $\eta_k(x)$ with complex time $x^0=t$, $\mathcal L$ transforms as a scalar field and normalization imposes constraint $S[0,0,0]=0$. Here $A_k$ denotes Lorentz scalars and $B^\mu_k$ -- fourvectors with the dot $\cdot$ denoting Minkowski scalar product.
In our convention, assuming that $\mathcal L$ is a scalar we also assumed that average charge densities are zero.
The first Minkowski coordinate (time) has the jump $x^0_i-x^0_f=i\beta$.

We stress that the charge conservation, Ward-Takahashi identity, holds in CTP if the Lagrangian does not mix
different classes of fermions \cite{peskin}
\be
\langle\partial_\mu j^\mu_N\cdots\rangle=0, \:j_N^\mu=\sum_{n\in N}\bar{\psi}_n\gamma^\mu\psi_n.
\ee

The great advantage of Lagrangian and path-integral formalism is that it seems to be perfectly Lorentz-invariant at first sight.
However, the possible caveats are hidden in the shape of CTP, which is bent into imaginary time but not spatial coordinates.
The common sense intuition tells us that the Lorentz invariance must be broken by finite temperatures, which essentially dictate
the contour jump. Moreover, even at zero temperature, the contour retains preferred time direction so
Lorentz invariance is not manifest and self-evident.

\subsection{Formal proof of invariance}

\begin{figure}
\includegraphics[scale=.5]{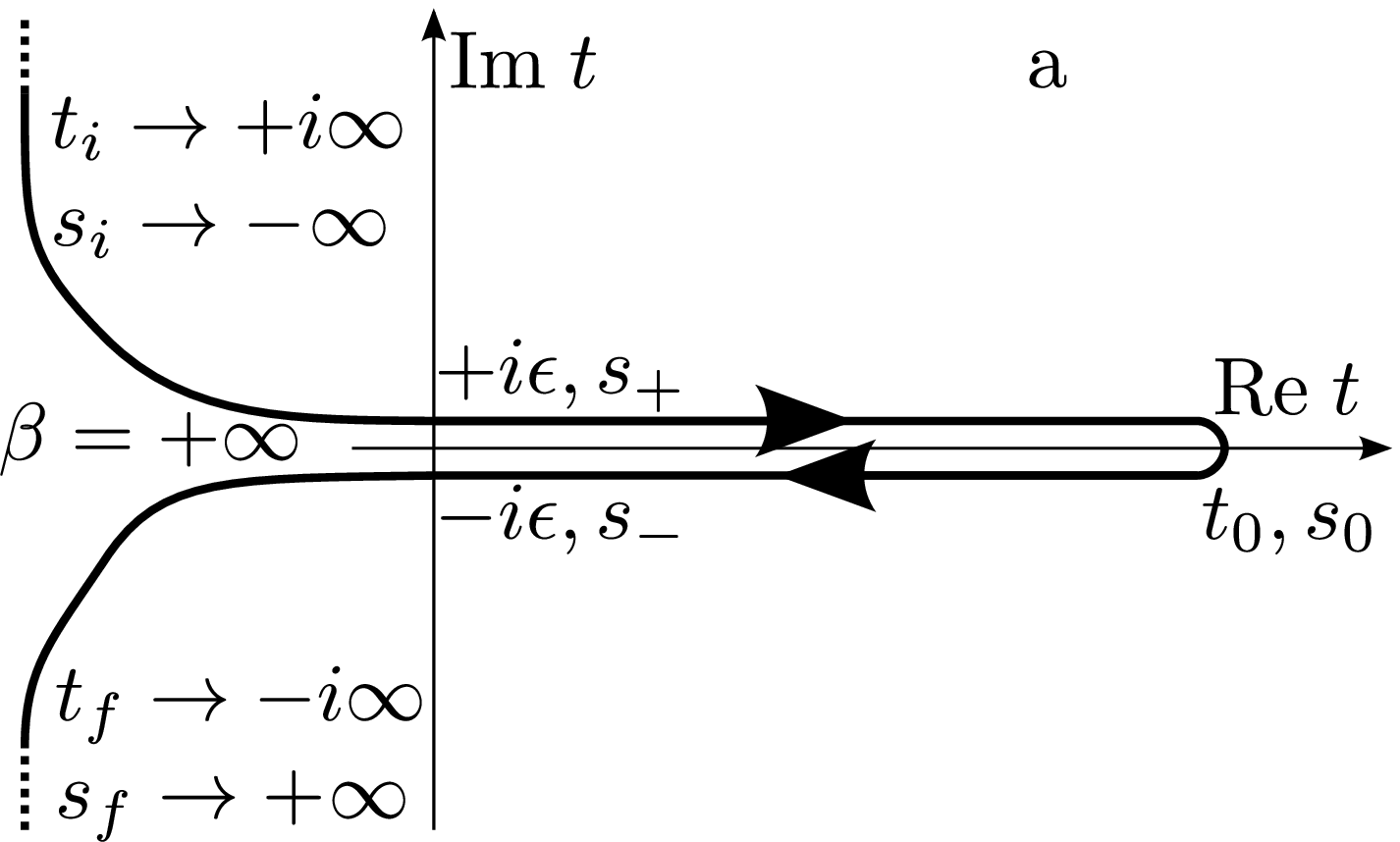}
\includegraphics[scale=.5]{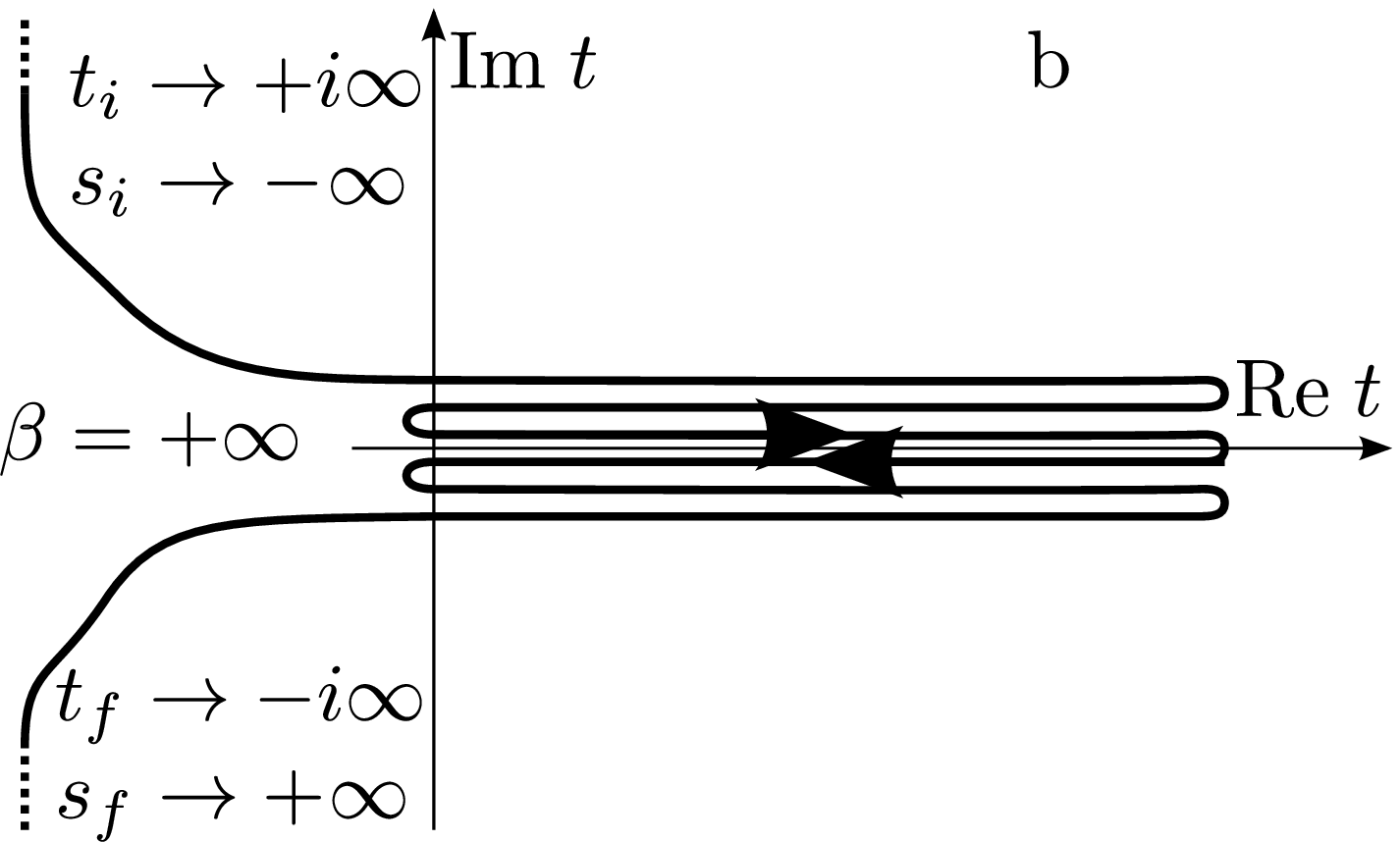}
\caption{Zero-temperature CTP. The boundary times $t_i$ and $t_f$ are sent to imaginary infinity, $i\infty$ and $-i\infty$, respectively. Similarly to Fig.\ref{ctppcon} we can take the simple contour (a) and the generalized one (b)}
\label{ctpz}
\end{figure}

We shall now present a brief formal proof of Lorentz invariance of vacuum generating function (\ref{genn}), with
the auxiliary field nonzero only for real times (right hand part of the contour) at zero temperature $k_BT=0$ or $\beta=+\infty$.
 Note that a zero temperature makes the jump on CTP between $t_i$ and $t_f$ infinite.
Due to the freedom of choice of the shape, we assume the contour in Fig. \ref{ctppcon} with $t_i\to i\infty$ and $t_f\to -i\infty$, see Fig. \ref{ctpz}.
Applying a boost $\Lambda$ to the fields $A\to \Lambda A$, $B\to \Lambda B$, $\psi\to \Lambda\psi$ we have the complex-time
invariance of Lagrangian
\ba
&&\int_C \rmd^4x\mathcal L[A,B,\psi]\to \int \rmd^4x\mathcal L(\Lambda A,\Lambda B,\Lambda\psi)=\nonumber\\
&&
\int_C \rmd^4x\Lambda_0\mathcal L[A,B,\psi]=\int_C \rmd^4x\mathcal L[A,B,\psi]
\ea
where we have used the fact that both the measure $d^4x$ and the integration boundary is invariant under Lorentz transformations and $\Lambda_0 f(x)=f(\Lambda^{-1}x)$.
Moreover \newline $D(\Lambda A)=DA$, $D(\Lambda B)=DB$ and $D(\Lambda\psi)=D\psi$.
 All the exponent in the generating functional (\ref{genl}) is a Lorentz scalar so the above reasoning remains valid. We write
 \ba
&&\int DADBD\bar\psi D\psi\; \exp\int_C \rmi\rmd^4x\left(\mathcal L[A,B,\psi]+\vphantom{\sum_k}\right.\\
&&\left.\sum_k(\Lambda\chi_kA_k+\Lambda\xi_k\cdot B_k+\Lambda\bar{\eta}_k\psi_k+\bar{\psi}_k\Lambda\eta_k)\right)\nonumber\\
&&=\int D(\Lambda A)D(\Lambda B)D(\Lambda \bar\psi)D(\Lambda \psi)\exp\int_C \rmi\rmd^4x\times\nonumber\\
&&\left(\mathcal L[\Lambda A,\Lambda B,\Lambda\psi]+\vphantom{\sum_k}\right.\nonumber\\
&&\left.\sum_k(\Lambda\chi_kA_k+\Lambda\xi_k\cdot B_k+\Lambda\bar{\eta}_k\Lambda\psi_k
+\Lambda\bar{\psi}_k\Lambda\eta_k)\right)=\nonumber\\
&&\int DADBD\bar{\psi}D\psi\; \exp\int_C \rmi\rmd^4x\times\nonumber\\
&&\left(\mathcal L[ A,B,\psi]+\sum_k(\chi_kA_k+\xi_k\cdot B_k+\bar{\eta}_k\psi_k
+\bar{\psi}_k\eta_k)\right),\nonumber
\ea
which proves the invariance. However, we had to assume that $\Lambda$ acts on functions of complex arguments, not just real domain, as time
is complex. To be correct we should work only in the real domain $(s,x^1,x^2,x^3)$ and define $\Lambda$ by Taylor expansion of
generator exponential
\be
\Lambda=\rme^{\rmi\omega_{\mu\nu}J^{\mu\nu}}=\sum_{n=0}^\infty \frac{1}{n!}(\rmi\omega_{\mu\nu}J^{\mu\nu})^n
\ee
This is equivalent to real Lorentz transformation for real times, which is the case we need (Green functions
at real times).
The above proof, although formally correct, is hardly acceptable because one uses ill-defined mathematical operations.
The vague issues are:
\begin{itemize}
\item Lorentz transformations are analytically continued to complex spacetime.
\item We ignored possible problems for $x^0\to\pm \rmi\infty$.
\item Convergence and renormalization is not discussed.
\end{itemize}

To give more insight to the proof, dispel possible objections and make it robust, we will now reformulate it in terms of perturbative diagrams
showing that each diagram separately is Lorentz-invariant.

It is enough to prove that $JG=0$ for all Lorentz generators $J$ and all Green functions $G$ -- defined by (\ref{gree}) at real times, because every finite continuous transformation can be written as $\Lambda=\exp wJ$ and $(\rmd/\rmd w)\Lambda G=J\Lambda G=JG_w$ where $\Lambda G=G_w$ is always some Green function at real times.

\subsection{Free theories}

We shall first prove Lorentz invariance of vacuum zero-temperature Green functions in free theories, where they can
be found exactly. The Lagrangian density for free scalar theory reads
\be
2\mathcal L_0=\partial_\mu A\partial^\mu A-m^2A^2,
\ee
where $m$ is the mass of the field.
The Green function (propagator) satisfies Klein-Gordon equation
\be
(\partial_\mu\partial^\mu+m^2)\langle A(x)A(x')\rangle_0=-\rmi\delta^4(x-x'),
\ee
where $\partial$ acts on $x$ (we shall write $\partial'$ if acting on $x'$).
The solution reads 
\be
G_0(x,x')=\langle A(x)A(x')\rangle_0=\label{gbos}\int \frac{\rmd^3p}{2p^0(2\pi)^3}\sum_\pm\mp\frac{e^{\pm \rmi p^\mu |x-x'|_\mu}}
{1-\rme^{\pm \beta p^0}}
\ee
where $p^\mu$ is the four momentum with $p^0=\sqrt{{\mathbf p}^2+m^2}$.
 It is important that (\ref{gbos}) is valid for time components, $x^0$ and $x^{\prime 0}$
 along the CTP contour as defined in Fig. \ref{ctppcon} and (\ref{x00}).
For the Dirac field,
\be
\mathcal L_0=\bar{X}(\rmi\gamma^\mu\partial_\mu -m)X
\ee
we have
\be
(-i\gamma^\mu\partial_\mu+m)_{bc}\langle X_c(x)\bar{X}_a(x')\rangle_0=-\rmi\delta_{ab}\delta^4(x-x'),
\ee
with the summation over $c$ and $m_{ab}=m\delta_{ab}$.
Now, we get two cases. 
If $X=A$ (bosonic field, unphysical, usually appearing in renormalization) then $A(t_i)=A(t_f)$ but if $X=\psi$ (fermionic field) then $\psi(t_i)=-\psi(t_f)$.
The general solution reads
\be
\langle X_a(x)\bar{X}_b(x')\rangle_0=(\rmi\gamma^\mu\partial_\mu+m)_{ab}G^X_0(x,x'),
\ee
where
\ba
&&G^A_0(x,x')=G_0(x,x'),\:\label{gfer}G^\psi_0(x,x')=\nonumber\\
&&\int \frac{\rmd^3p}{2p^0(2\pi)^3}\sum_\pm\mp\frac{\rme^{\pm \rmi p^\mu |x-x'|_\mu}}
{1+\rme^{\pm \beta p^0}}.
\ea
For vector fields $B^\mu$, the internal Lorentz structure is simply added by metric tensor \newline $\langle B^\mu(x)B^\nu(x')\rangle_0=ag^{\mu\nu}G_0(x,x')$.
Possible problems occur for expressions of the type $\partial_\mu\partial'_\nu G_0$ because the derivative produces unphysical
contact terms $\sim \delta(x^0-x^{\prime 0})$ \cite{itzyk}. Fortunately, they appear always in pairs with a renormalization field, which cancels the singularity.

Nonzero temperature $\beta<\infty$ will break Lorentz invariance, which means that
there is a four-vector related to temperature. There is no unique choice. One can define
$\beta^\mu$  \cite{furnser,weld} with $\beta^0=\beta$ and $\beta^k=0$, $k=1,2,3$ in a particular reference frame.
However, good alternatives are $k_BT^\mu=\beta^\mu/\beta\cdot\beta$
or $u^\mu=\beta^\mu/\sqrt{\beta\cdot\beta}$, see also a general discussion \cite{hanggi}. However, one has to remember that the zero temperature limit means
\emph{infinitely time-like} $\beta^\mu$.  The zero four-vector $\beta$ limit, when Lorentz invariance might be obvious,
corresponds unfortunately to infinite temperature so this is not what we are looking for.

In the zero-temperature limit $\beta\to+\infty$, (\ref{gbos}) and (\ref{gfer}) take the form
\be
G^\psi_0(x,x')=G_0(x,x')=\int\frac{\rmd^3p}{2p^0(2\pi)^3}\rme^{- \rmi p^\mu |x-x'|_\mu}\label{gzero}
\ee
Applying Lorentz generator
\be
J_0=J_{0x}+J_{0x'},\;J^{\mu\nu}_{0x}=\rmi(x^\mu\partial^\nu-x^\nu\partial^\mu)
\ee 
acting on $x$ and $x'$ by $J_{0x}$ and $J_{0x'}$, respectively,
one can check explicitly (a lengthy textbook exercise) that $JG=0$ (so $\Lambda G=G$) when applied to (\ref{gzero}), so the free theory is indeed Lorentz-invariant
(there are no additional problems with transformation of internal vector or spinor structure).
Remember that the time derivative in $J$ is always taken along the contour, namely
with complex-time derivative $\partial_0$ (\ref{timeim}).

\subsection{Interaction}

The rest of the proof is similar to the shape-independence proof. 
Again, it is important that all time components below, $x^0_k$, run
 along the CTP contour as defined in Fig. \ref{ctppcon} (in particular Fig. \ref{ctpz} at zero temperature) and (\ref{x00}). 
The Green functions 
\be
\left\langle \prod_{k} X_k(x_k)\right\rangle
\ee
are, by virtue of the Wick theorem,
represented by sums of diagrams with external vertices $k$ (not integrated), internal vertices (integrated) corresponding to the interaction part of the Lagrangian density $\mathcal L_I$ and connecting lines
corresponding to free Green functions (\ref{gbos}) and (\ref{gfer}).
We want to show that $J$ vanishes when applied to the whole diagram. Suppose that the diagram has fixed number $N$ of internal vertices $1..N$ and $K$ external ones $N+1..N+K$.
then we have to show that
\be
\int \rmd^4x_1\cdots \rmd^4x_N\sum_{l>N} J^l\left\langle\prod_{n\leq N}\mathcal L_I(x_n)\prod_{k>N} X_k(x_k)\right\rangle_0=0,\label{jj0}
\ee
where $X_k$ represent fields whose expectation value define the Green function, $J^m$ acts on all fields meeting at the  vertex $m$. The average is taken with respect to free Lagrangian density $\mathcal L_0$. 
We decouple (\ref{jj0}) into free Green functions using Wick theorem and use Leibniz rule. Since Lorentz transformations
do not affect free Green functions, we have
\be
\sum_{m=1}^{N+K} J^m\left\langle\prod_{n\leq N}\mathcal L_I(x_n)\prod_{k>N} X_k(x_k)\right\rangle_0=0.
\ee
Since $\mathcal L_I$ is a Lorentz scalar we can change $\sum_nJ^n\to\sum_n J^n_0$. Finally, we perform integration 
\be
\int d^4x\, J^{\mu\nu}_0\cdots\label{jjj}
\ee
over each vertex $n\leq N$ ($x=x_n$),
which yields only boundary values due to $\int \rmd t\;x^k\partial_0=\int \rmd s\;x^k\partial_s$, see Eq. (\ref{j00}). However, free propagators vanish exponentially at large distances and, at zero temperature, with $x^{0}\to\pm i\infty$.
Hence (\ref{jjj}) vanishes along with the left hand side of (\ref{jj0}), which completes the proof.

This is our main result. The proof has been obtained in the time domain, without stretching the real time
to infinities. Note the analogy to contour shape independence. The heart of the proof is the fact that
analytic continuation of time does not hurt the generating function. It is quite intuitive but
never before shown explicitly.

\section{Incomplete proofs}

It is tempting to ask:
Why not to prove Lorentz invariance using simply energy-momentum representation
\cite{niemi,kobes2,landweert}?
This approach seems attractive and practical but it has to be supplemented by arguments
following from spacetime CTP representation. The main problems arise for off-shell Green functions (not describing scattering processes)
and field mass/strength renormalization. They cannot be resolved without returning to spacetime representation.

The energy-momentum representation can be used only on the real part of the contour 
stretched to the whole real axis ($t_0\to +\infty$ and $t(s_\pm)\to -\infty$), taking separately its upper and lower branch,
as Fourier transform is only defined in real axis. The Green function gets additional indices to keep track of the branch,
$G_{uv}(x,x')=G(x_u,x'_v)$ where $u,v=\pm$ and $x^0_\pm=x^0\pm \rmi\epsilon$, $\epsilon\to 0_+$.
We have
\be
G(x,x')=\int\frac{\rmd^4p}{(2\pi)^4}G(p)\rme^{-\rmi p^\mu (x_\mu-x'_\mu)},
\ee
where, in the space $(+,-)$,
\ba
&&G^{\pm}(p)=\left(\begin{array}{cc}
\rmi(p^2-m^2+\rmi\varepsilon)^{-1}&2\pi\delta_-(p^2-m^2)\\
2\pi\delta_+(p^2-m^2)&-\rmi(p^2-m^2-\rmi\varepsilon)^{-1}\end{array}\right)\nonumber\\
&&\pm 2\pi\delta(p^2-m^2)n^\pm(|p^0|)\left(\begin{array}{cc}
1&1\\
1&1\end{array}\right),
\ea
where $\delta_\pm(p^2-m^2)=\theta(\pm p^0)\delta(p^2-m^2)$, the upper index $+/-$ denotes bosonic/fermionic field with
$n^\pm(q)=(e^{\beta q}\mp 1)^{-1}$. We denoted $\varepsilon\to 0_+$ defined for momenta to distinguish it from $\epsilon$ defined
in spacetime.

\begin{figure}
\includegraphics[scale=.5]{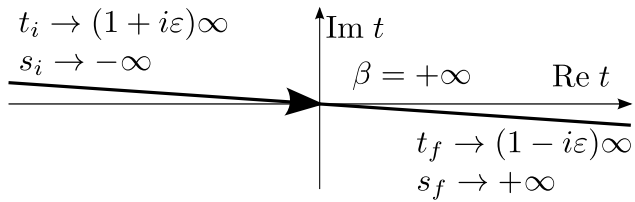}
\caption{The tilted contour which circumvents the problem of pole renormalization $p^2\sim m^2\to M^2$}\label{til}
\end{figure}

In the zero-temperature limit $n\to 0$ and the free Green functions are manifestly Lorentz-invariant. The problems arise for interactions.
Now, the interacting Green function takes on renormalization corrections,
\be
G_{0++}(p)=\frac{\rmi}{p^2-m^2+\rmi\varepsilon}\to G_{++}(p)=\frac{\rmi Z(p)}{p^2-M^2+\rmi\varepsilon},
\ee
where $Z(p)$ is regular at $p^2=M^2$. The renormalization is obtained by summation of Dyson series of self-energy contributions.
However, such step is ill-defined for $p^2=M^2$ because $\varepsilon$ dominates in this regime. This can be cured
by a slight tilt of the real-time contour as in Fig. \ref{til} \cite{peskin}. In fact, the new contour contains the essential feature of Fig. \ref{ctpz} -- the
endpoints in $\pm i\infty$. However, for $G_{\pm\mp}$ the trick with the contour does not help. We are forced to 
simply accept
\be
\delta_\pm(p^2-m^2)\to Z\delta_\pm(p^2-M^2)
\ee
because this part is only defined for $p^2=M^2$. One can circumvent the problem by \emph{ad hoc} defining extrapolation of $G_{\pm\pm}$
to $G_{\pm\mp}$ and the $p^2=M^2$ case. Another argument is the fluctuation dissipation theorem, which imposes relations
between elements of $G$ \cite{fdt,chou}, or mass derivative \cite{nieg}. However, this is only the case of elementary
particles like electron and photon. Composed bound states like nucleons, atoms, and molecules get their renormalization
corrections, too, but they cannot be simply reduced to a single dressed propagator line. In the time domain, bound
states do not lead to any problems because the poles/deltas in propagators appear only in the infinite time limit
and we do not have to sum infinite series of diagrams.

\begin{figure}
\includegraphics[scale=.5]{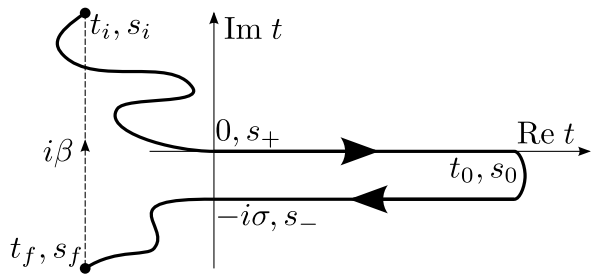}
\caption{The separated upper and lower real contour branches $0\leq\sigma\leq \beta$}\label{ctpg}
\end{figure}

\begin{figure}
\includegraphics[scale=.5]{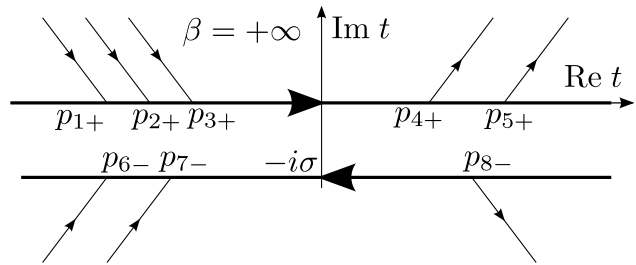}
\caption{The particle view of factorization mechanism. The Green function is $\sigma$-independent only if $p^0_{1+}+p^0_{2+}+p^0_{3+}+p^0_{4+}+p^0_{5+}=0$
and $p^0_{6-}+p^0_{7-}+p^0_{8-}=0$}\label{ctppm}
\end{figure}

Further problems may arise from pinch singularities  \cite{jackiw}, which may occur for products of propagators $[(p^2-m^2+\rmi\varepsilon)(p^2-m^2-\rmi\varepsilon)]^{-1}$.
Fortunately, a careful analysis makes them always cancel \cite{kobes2}. Having the invariance of in-out correlations, it would be easy to prove Lorentz invariance if all the Green functions factorized at zero temperature
\cite{landweert,furnser} , namely
\be
\left\langle\prod X_k(p_k)\right\rangle=\left\langle\prod_{k+} X_{k}(p_k)\right\rangle\left\langle \prod_{k-} X_k(p_k)\right\rangle\label{fact}
\ee
where $k\pm$ denotes fields defined on the upper($+$)/lower($-$) branch by the Fourier transform\newline $X_{\pm}(p)=\int \rmd^4x_\pm \rme^{\rmi p^\mu x_{\mu\pm}}X_{\pm}(x_\pm)$.
An attempt to prove the factorization requires generalization of CTP to the contours like that in Fig. \ref{ctpg} with varying $\sigma$ \cite{niemi,landweert,furnser}. If $\sigma=\beta/2$ then the 
factorization occurs indeed for $\beta\to\infty$. However, not every Green function is $\sigma$-independent \cite{evan2,nieg06}. From translational invariance
the nonvanishing Green functions satisfy $\sum p=0$. The new vertical part of the contour gives additional factors $\rme^{\pm\sigma p^0}$ to each propagator
from lower to upper branch ($+$) or vice versa ($-$). These factors cancel only if
\be
\sum_{k+}p^0_k=\sum_{k-}p^0_k=0,
\ee
which means that the frequencies/energies of upper and lower branch \emph{separately} add up to zero as seen in 
Fig. \ref{ctppm}. This situation is common to scattering processes but not in general. 
It holds e.g. for in-out correlations when all times are on the upper real part due to time shift invariance but
not necessarily for in-in correlations when times are located on both real parts (upper and lower).

Having presented above the complications in energy-momentum space, we conclude that the proof is fully correct in the spacetime domain, while in the
momentum/Fourier space it would require assuming at best several additional technical rules and
may not be general (for very complicated diagrams). 
Still, most of practical calculations are easier performed in the latter
case, with spacetime arguments used only when the momentum rules are ambiguous.

Another attempt of proof requires reconstruction of vacuum wavefunction out of in-out Green functions. This is possible in
simple quantum models, including free theories, finite and harmonic systems but not in fully interacting relativistic theory because renormalizing fields, especially for fermions, make the information in in-out functions incomplete and insufficient to uniquely reconstruct vacuum wavefunction. On the other hand, dimensional regularization is useless when
proving Lorentz invariance because Lorentz transformations are defined only for integer dimensions.
Lastly, an often heard argument is analyticity of Green functions \cite{nair}, but how to define it on CTP? The condition
of nonincreasing imaginary part makes it impossible to define analytic conditions (Cauchy-Riemann) because for
free real times the middle one is pinched -- the imaginary part of the derivative is not allowed. 
One cannot allow the contour to go upwards in imaginary direction because the energy spectrum is bounded from below and open form above, making integrals nonrenormalizably divergent. 
On the other hand analyticity with respect to real $s$ is questionable because of turning points $t_0(s_0)$
and crossings ($t_n\simeq t_m$). Such a proof will be restricted to a particular branch of CTP, e.g. upper/lower real or vertical imaginary part, but not general.
One cannot also use Gell-Mann and Low theorem \cite{gem}, which  could simply extend the proof from free theories to interacting ones by adiabatic switching, because it works itself in a preferred frame.
Any attempt to prove Lorentz invariance in general will ultimately fail or arrive at the proof here presented.

\section{Absence of time-reversal symmetry}

The momentum representation hides the fact that the Lorentz invariance is not necessarily valid for time reversal, which is only
true for second order correlations or if the factorization (\ref{fact}) holds for higher order correlations. Note that the reversal $p^0\to -p^0$
break the time symmetry because of $\theta(\pm p^0)$ factors in $G_{\pm\mp}$. Trying to reverse branches $+\leftrightarrow -$ does not help either,
because then $G_{\pm\pm}$ gets conjugated, so there is no way to restore original Green function. If we both reverse $p^0$ and make conjugation then
we will return to the original Green function but the time will be also twice reversed which is again not what we wanted to have. The time (generally charge-parity-time) symmetry is valid only for scattering processes 
or space-like separated points. This lack of symmetry is fundamental and especially surprising in context of noninvasive measurements which
are naturally defined on CTP \cite{our}.

\section{Conclusions}

We have shown that the closed time path formalism is consistent with relativistic invariance at zero temperature.
The presented picture path-integral Lagrangian fits well into
covariant relativistic framework with the analytic continuation of time as the only nontrivial extra element.
It is remarkable that complex time glues statistical mechanics, time evolution and relativistic symmetry
without any problems.
The fact that CTP is invariant only under continuous transformations, not time reversal, shows that there
the properties of CTP still differ from scattering processes. An open issue is whether the Lorentz symmetry
can be shown nonperturbatively (what about spontaneous breaking?) and for quantum gravity.

\section*{Acknowledgements}

I thank W. Belzig, A. Neumaier, S. Mr\'owczy\'nski and P. Chankowski for fruitful discussions and motivation
and for hospitality of University of Konstanz, where part of this work has been completed.

\appendix
\section{Hamiltonian formulation of CTP}
\label{appa}
\renewcommand{\theequation}{A.\arabic{equation}}
\setcounter{equation}{0}
\renewcommand{\thesubsection}{(\alph{subsection})}
\setcounter{subsection}{0}
The traditional starting point of every quantum-mechanical problem is the Hermitian Hamiltonian operator $\hat{H}$ in a given Hilbert space. The physical quantities are described by Hermitian operators $\hat{X}_k$ corresponding to
their classical counterparts $X_k$ (numbers). Depending on the measurement scheme, there exists a correspondence
between operators and measurement outcomes, $\hat{X}_k\to X_k$. Evolution and state is described by Hamiltonian $\hat{H}(t)\to\hat{H}+\hat{H}_1(t)$, where $\hat{H}_{1}(t)$
is the time dependent part, due to changing external forces, which are assumed to be absent for $\mathrm{Re}\,t<0$. 
We shall assume that $t$ lies on CTP as in Fig. \ref{ctppcon}.
Every conserved quantity
can be incorporated into the definition of the Hamiltonian, $\hat{H}\to\hat{H}-\mu_Q\hat{Q}$.  For the main purpose of this paper, $\hat{H}_1$ and $\hat{Q}$ can be omitted. 
However, the whole CTP formalism works perfectly
also for nonzero $\hat{H}_1$, which is necessary in problems involving external (changing) forces and charges \cite{kame}. 

The generating functional can be written in the form
\be
\rme^{S[\chi]}=\frac{\mathrm{Tr}\mathcal T_Ce^{\int_C \rmi\rmd t\left(\sum_k\chi_k(t)\hat{X}_k-\hat{H}(t)\right)}}
{\mathrm{Tr}\mathcal T_Ce^{-\int_C \rmi \rmd t\hat{H}(t)}}\label{gen1},
\ee
where the subscript $C$ denotes integration over the CTP as defined in Sec. \ref{ctps} and the
time order  $\mathcal T_C$ denotes  ordering of operators in the Taylor expansion according to CTP, namely
\be
\mathcal T_C\hat{Y}_1(t_1)\cdots \hat{Y}_n(t_n)=\mathrm{sgn}(\sigma_f)\hat{Y}_{\sigma(1)}(t_{\sigma(1)})\cdots\hat{Y}_{\sigma(n)}(t_{\sigma(n)}),
\label{ord}
\ee
where $\sigma$ is a permutation such that $s_{\sigma(1)}\geq \dots\geq s_{\sigma(n)}$ (in the case of $\hat{X}_k$ we take the time of $\chi$). Here $\mathrm{sgn}(\sigma_f)$ denotes
the sign of the sub-permutation  fermionic operators. If $\hat{Y}_k$ is a fermionic operator for $k\in\{1..n_f\}$,
then $\sigma_f$ is defined as the map between the ordered sequences $(\sigma(k))_{k=1}^{n_f}$ and $(\sigma(k))$,
ordered according to $k$ and $\sigma(k)$, respectively.
 
The correlations (Green functions) are given by derivatives of the generating functional,
\be
\left\langle\prod_k X_k(t_k)\right\rangle_{q,\rho}=\prod_{k}\frac{-\rmi\delta}{\delta \chi_k(t_k)}
\left.\rme^{S[\chi]}\right|_{\chi=0}.\label{avgs}
\ee 

Performing a Taylor expansion of exponentials in (\ref{gen1}) one can explicitly check that the result does not depend on a particular shape of CTP \cite{gelis},
as long as (\ref{ctpd}) is satisfied and $t_0$ is greater than all interesting times. In such a check it is important to
note that $\hat{H}$ is time-independent along the wiggly part of the contour, so the ordering does not matter there.
On the other hand (\ref{gen1}) is useful also for deriving the
thermodynamic functions and transport coefficients, e.g. the Kubo formula \cite{kms}.

Usually one separates free harmonic Hamiltonian $\hat{H}_0$ writing $\hat{H}(t)=\hat{H}_0+\hat{H}_I(t)$.
The remaining calculations are usually performed perturbatively, with the help of the free two-point Green's functions and Wick's theorem \cite{wick,wickt}, which allows a decoupling of the free many-point Green functions into products of two-point functions, see below.

\subsection{Free Green functions}

The most convenient free Hamiltonian is the quadratic form of bosonic and fermionic operators
\be
\hat{H}_0=\sum_{kl}(b_{kl}\hat{x}_k\hat{x}_l+f_{kl}\hat{\phi}_k\hat{\phi}_l)
\ee
where $\hat{x}$ and $\hat{\phi}$ are Hermitian operators
with bosonic and fermionic commutation relations, respectively,
$[\hat{x}_k,\hat{x}_l]=\rmi g_{kl}\hat{1}$, $\{\phi_k,\phi_l\}=h_{kl}\hat{1}$, with real $g$ and $h$.

One can always diagonalize $\hat{H}_0$ so that
\be
\hat{H}_0=H_{vac}+\sum_{k}E_k(\hat{A}^\dag_k\hat{A}_k+\hat{\psi}^\dag_k\hat{\psi}_k)
\ee
where $H_{vac}$ is the vacuum energy (can be ignored), $\hat{A}$ and $\hat{\psi}$ are linear combinations of $\hat{x}$ and $\hat{\phi}$, respectively,
with the property 
\be
[\hat{A}_k,\hat{A}_l]=0,\:[\hat{A}_k,\hat{A}^\dag_l]=\delta_{kl},\:\{\hat{\psi}_k,\hat{\psi}_l\}=0,\:\{\hat{\psi}_k,\hat{\psi}^\dag_l\}=\delta_{kl}.
\ee
It is especially simple and instructive to find Green functions for the above Hamiltonian, according to (\ref{avgs}) and (\ref{gen1}), extended to the whole CTP.
In the case of bosonic operators, $\langle A(t)\rangle_0=\langle A^\ast(t)\rangle_0=0$ and
\begin{eqnarray} 
&&\langle A_k(t)A_l(t')\rangle_0=\langle A^\ast_k(t)A^\ast_l(t')\rangle_0=0,\label{bgreen}\\
&&\langle A_k(t)A^\ast_l(t')\rangle_0=\delta_{kl}\rme^{\rmi(t'-t)E_k}\left[\frac{\theta(t-t')}{1-\rme^{-\beta E_k}}+\frac{\theta(t'-t)}{\rme^{\beta E_k}-1}\right].\nonumber
\end{eqnarray}

For fermionic operators, it is important that $\phi$ ($\psi$, $\psi^\ast$) is never a $c$-number but a Grassmann number 
with the property $\phi_a\phi_b=-\phi_b\phi_a$, which follows from anticommutation relations.
The anticommutation makes it necessary to use different boundary conditions, due to Kubo, Martin and Schwinger \cite{kms}, imposing the sign reversal
 on the jump $t_f\to t_i$, so the Green functions read $\langle \psi(t)\rangle_0=\langle\psi^\ast(t)\rangle_0=0$ and
\begin{eqnarray}
&&\langle \psi_k(t)\psi_l(t')\rangle_0=\langle \psi^\ast(t)_k\psi^\ast_l(t')\rangle_0=0,\nonumber\\
&&\langle \psi_k(t)\psi^\ast_l(t')\rangle_0=\delta_{kl}\rme^{\rmi(t'-t)E_k}\times\nonumber\\
&&\left[\frac{\theta(t-t')}{1+\rme^{-\beta E_k}}-\frac{\theta(t'-t)}{\rme^{\beta E_k}+1}\right].\label{fgreen}
\end{eqnarray}
In both cases, the Green functions satisfy the equation
\ba
&&\delta(t-t')\delta_{kl}=[\partial_t+\rmi E_k]\langle A_k(t)A^\ast_l(t')\rangle_0\nonumber\\
&&=[\partial_t+\rmi E_k]\langle\psi_k(t)\psi^\ast_l(t')\rangle_0\label{egreen}
\ea
with different boundary conditions, $A(t_i)=A(t_f)$ but $\psi(t_i)=-\psi(t_f)$.

\subsection{Wick theorem}

The many-point Green functions are obtained from Wick theorem \cite{wick,wickt}. For products of odd number of operators,
Green function vanishes while for the even number $2n$,
\ba
&&\left\langle \prod_{k=1}^{2n} x_k(t_k)\right\rangle_0=\nonumber\\
&&\frac{1}{n!2^n}\sum_\sigma
\prod_{k=1}^n\langle x_{\sigma(2k-1)}(t_{\sigma(2k-1)})x_{\sigma(2k)}(t_{\sigma(2k)})\rangle_0,\nonumber\\
&&\left\langle \prod_{k=1}^{2n} \phi_k(t_k)\right\rangle_0=\\
&&\frac{\mathrm{sgn}\,\sigma}{n!2^n}\sum_\sigma
\prod_{k=1}^n\langle \phi_{\sigma(2k-1)}(t_{\sigma(2k-1)})\phi_{\sigma(2k)}(t_{\sigma(2k)})\rangle_0\nonumber
\ea

The Wick theorem states essentially that every many-point Green function for a quadratic Hamiltonian splits into
products of two-point functions. This fact is analogous to the property of Fourier transforms of Gaussian functions,
which are again Gaussian. Wick theorem is useful for interactions -- one proceeds perturbatively, expanding (\ref{gen1}) in interaction strength.

\subsection{Hamiltonian-Lagrangian equivalence for bosonic fields}

Let us begin with the Lagrangian
\be
L(q,\dot{q},t)=\sum_k(m_k\dot{q}_k^2/2+f_k(q)\dot{q}_k)-g(q)
\ee
for generalized coordinates $q$ and $\dot{q}=dq/dt$.
The classical Hamiltonian is obtained by Legendre transformation, defining momenta,
\be
p_k=\partial L/\partial \dot{q}_k=m_k\dot{q}_k+f_k(q)
\ee
so that the Hamiltonian reads
\be
H(q,p)=\sum_k p_k\dot{q}_k-L=\sum_k(p_k-f_k(q))^2/2m_k+g(q).
\ee
We assume quantum evolution of the wavefunction $\Psi(q)$ given by the path integral
\be
\Psi(\tilde{q},t)=\int_{q(0)=q'}^{q(t)=\tilde{q}} Dq\;\rme^{\int \rmi L(q,\dot{q})\rmd t}\Psi(q',0),
\ee
where $Dq$ includes all necessary normalization factors.
To find the operator form of the Hamiltonian, we expand the above equation for small times,
\ba
&&\Psi(\tilde{q},t)=\int^{q(t)=\tilde{q}}Dq\;\rme^{\int_0^t\rmi\rmd t'\sum_km_k\dot{q}_k^2(t')/2}\times\nonumber\\
&&\left(1-g(\tilde{q})t+\sum_k\int_0^t\rmi\rmd t'\dot{q}_k(t')f_k(\tilde{q})\right.\nonumber\\
&&\left.-\int_0^t\rmd t'\dot{q}_k(t')\frac{\partial}{\partial\tilde{q}_k}
+\sum_{kl}\int_0^t\rmd t'\rmd t''\dot{q}_k(t')\dot{q}_l(t'')\right.\label{expan}\\
&&\left(\frac{\partial^2}{2\partial\tilde{q}_k\partial\tilde{q}_l}-i\partial f_k(\tilde{q})\frac{\partial}{\partial\tilde{q}_l}
-f_k(\tilde{q})f_l(\tilde{q})/2\right)\nonumber\\
&&\left.-\sum_{kl}\int_0^t\rmi \rmd t'\dot{q}_k(t')\int_0^{t'}dt''\dot{q}_l(t'')\frac{\partial f_k(\tilde{q})}{\partial\tilde{q}_l}\right)\Psi(\tilde{q},0).\nonumber
\ea
The ignored terms are of the order $\sim t^2$. We now use the fact that
\ba
&&\int Dq\,\rme^{\int \rmi\rmd t'\sum_km_k\dot{q}_k^2(t')/2}\dot{q}_k(t')=0,\\
&&\int Dq\,\rme^{\int \rmi\rmd t'\sum_km_k\dot{q}_k^2(t')/2}\dot{q}_k(t')\dot{q}_l(t'')=\frac{\rmi}{m_k}\delta_{kl}\delta(t'-t''),\nonumber
\ea
which gives
\ba
&&(\Psi(\tilde{q},t)-\Psi(\tilde{q},0))/\rmi t=\left(-g(\tilde{q})
+\sum_k\frac{1}{m_k}\times\right.\\
&&\left.\left(\frac{\partial^2}{2\partial\tilde{q}_k^2}-
f_k(\tilde{q})\frac{\rmi\partial}{\partial\tilde{q}_k}
-f_k^2(\tilde{q})/2
-\frac{\rmi\partial f_k(\tilde{q})}{2\partial\tilde{q}_k}\right)\right)\Psi(\tilde{q},0).\nonumber
\ea
Let us shortly comment the last term in bracket. It has been obtained by assuming $\theta(t)\delta(t)\to\delta(t)/2$
or symmetrizing the last integral in (\ref{expan}) \cite{zinn}.
However, in the fundamental theories, like quantum electrodynamics, weak, strong interactions and generally Standard Model this term is absent.
It follows from the fact that the even the most dangerous terms, due to non-Abelian gauge fields, have $\partial f_k/\partial q_k=0$.
Nevertheless, this indicates potential problems for completely general Lagrangians, especially for gravity.

Finally, we get
\be
\rmi\partial_t\Psi=\hat{H}\Psi,\:
\hat{H}=\sum_k(\hat{p}_k-f_k(q))^2/2m_k+g(q),\;\hat{p}_k=\frac{\partial}{\rmi\partial q_k},
\ee
which proves the equivalence between bosonic path integrals and Hamiltonian picture, with the commutation rule $[\hat{q}_k,\hat{p}_l]=\rmi\delta_{kl}$.

\subsection{Hamiltonian-Lagrangian equivalence for fermionic fields}

Fermionic path integrals are more complicated due to anticommutation rules. We have to define abstract anticommuting Grassmann numbers, $\phi_k$,
with the property $\phi_k\phi_l=-\phi_l\phi_k$ and the integrals \cite{peskin},
\be
\int \rmd\phi_k=0,\;\int \rmd\phi_k\,\phi_k=1,\:\int \rmd\phi_1\cdots \rmd\phi_n\,\phi_n\cdots\phi_1=1.
\ee
One can check that
\ba
&&\int D\phi^\ast D\phi \rme^{-\sum_{kl}\phi^\ast_kF_{kl}\phi_l}=\det F,\\
&&\int D\phi^\ast D\phi \rme^{-\sum_{kl}\phi^\ast_kF_{kl}\phi_l}\phi_j\phi^\ast_m=(F^{-1})_{jm}\det F.\nonumber
\ea
Here $\phi=\phi_r+\rmi\phi_i$ and $\phi^\ast=\phi_r-\rmi\phi_i$ are independent fields
and $\rmd\phi^\ast d\phi=\rmi\rmd\phi_i\rmd\phi_r/2$.

Let us now take an abstract Lagrangian
\be
L_0(\psi,\dot{\psi},t)=\sum_k\psi^\ast_k(i\dot{\psi}_k-E_k\psi_k).
\ee
According to the classical rule we can construct an abstract Hamiltonian
\be
H_0=\sum_k p_k\dot{\psi}_k-L_0=\sum_k E_k\pi_k\psi_k
\ee
for
\be
\pi_k=\frac{\partial L_0}{\partial\psi_k}=i\psi^\ast_k.
\ee

In the continuous case for CTP,
\ba
&&\frac{\int D\psi^\ast D\psi \rme^{\sum_n\int_C \rmi\rmd t\psi^\ast_n(\rmi\partial_t-E_n)\psi_n}\psi_k(\tilde{t})\psi^\ast_l(t')}
{\int D\psi^\ast D\psi e^{\sum_n\int_C \rmi\rmd t\psi^\ast_n(\rmi\partial_t-E_n)\psi_n}}\label{ppp}\\
&&=\delta_{kl}\rme^{\rmi(t'-\tilde{t})E_k}
\left[\frac{\theta(\tilde{t}-t')}{1+\rme^{-E_k/k_BT}}-\frac{\theta(t'-\tilde{t})}{\rme^{E_k/k_BT}+1}\right]
\nonumber
\ea
which is the equivalent to (\ref{fgreen}). The equivalence between bosonic-fermionic path integrals
and Hamiltonian operator can be now easily extended to general family of Lagrangians of the type
\be
L=\sum_k(m_k\dot{q}^2_k/2+\dot{q}_kf_k(q,\psi)+\rmi\psi^\ast_k\dot{\psi}_k)-g(q,\psi),\label{lag1}
\ee
where $f_k(q,\psi)$ and $g(q,\psi)$ are real-valued and even in $\psi$,
and Hamiltonian
\be
\hat{H}=\sum_k(\hat{p}_k-f_k(\hat{q},\hat{\psi}))^2/2m_k+g(\hat{q},\hat{\psi})
\ee
with commutation relations $[\hat{p}_k,\hat{q}_l]=i$ and $\{\hat{\psi}_k,\hat{\psi}^\ast_l\}=\delta_{kl}$.
We emphasize, however, that the equivalence does not generalize to completely arbitrary Lagrangians, containing higher powers of time
derivatives, e.g. $\dot{q}^4$ or $\dot{\psi}^2$.

\section{Technical aspects of path integrals}
\label{techa}
\renewcommand{\theequation}{B.\arabic{equation}}
\setcounter{equation}{0}
\renewcommand{\thesubsection}{(\alph{subsection})}
\setcounter{subsection}{0}
\subsection{Wick theorem for path integrals}
\label{wicka}
The Wick theorem is now a straightforward consequence of Gaussian integrals
\ba
&&\int DqD\phi \rme^{-\sum_{kl}(q_kg_{kl}q_l-\phi_kb_{kl}\phi_l)/2+\sum_k(\chi_kq_k+\eta_k\phi_k)}
\nonumber\\
&&\propto \rme^{\sum_{kl}(\chi_kg^{-1}_{kl}\chi_l-\eta_kb^{-1}_{kl}\eta_l)/2}.
\ea
where $g$ and $b$ are real symmetric and antisymmetric matrices, respectively, while $g^{-1}$ and $b^{-1}$ are their inverses.
\subsection{Conservation laws}
\label{consa}

As noticed already by Kadanoff and Baym \cite{kaba}, the CTP technique maintains classical conservation laws.
Suppose that the fermionic terms in Lagrangian (\ref{lag1}) appear in separate sets,
\be
L=\sum_{n\in N}\psi^\ast_ni\partial_t\psi_n+L_I,
\ee 
where $L_I$ is built from bosonic field and fermionic quadratic forms
\be
\sum_{N}\sum_{n,n'\in N}\psi_n^\ast\psi_{n'}.\label{kk1}
\ee
Sets $N$ form a family of \emph{disjoint} sets and $L_I$ may contains arbitrary products and sums
of (\ref{kk1}) for different $N$s together with bosonic fields. It is, however, forbidden to include terms
like $\psi^\ast_k\psi_{k'}$ where $k$ and $k'$ belong to different sets $N$ and $N'$, respectively.
Each set $N$ defines one conserved quantity, $Q_N(t)=\sum_{n\in N}\psi^\ast_n(t)\psi_n(t)$.
The conservation law states that
\be
\langle \partial_t Q_N(t)\cdots\rangle=0,\label{conse}
\ee
when averaging with the path integral (\ref{genn}). To prove (\ref{conse}) we make the transformation
$\psi_n(t)=e^{i\alpha(t)}\psi'_n(t)$, for real $\alpha$ with boundary condition $\alpha(t_i)=\alpha(t_f)$.
As the transformation is linear, $D\psi^\ast D\psi=D\psi^{\prime\ast}D\psi'$ and
\ba
&&DAD\psi^\ast D\psi\,\rme^{\int_C idt\,L}=DAD\psi^\ast D\psi\,e^{\int_C \rmi\rmd t\,\left(L(t)-Q_N\partial_t\alpha\right)}=\nonumber\\
&&DAD\psi^\ast D\psi\,\rme^{\int_C \rmi\rmd t\,\left(L(t)+\alpha\partial_t Q_N\right)}
\ea
where we performed integration by parts with respect to time in the last line. Taking the above equation integrated with arbitrary $Q_N$-conserving field function
and preforming the functional derivative
\be
\delta/\delta\alpha(t)|_{\alpha=0}
\ee
we obtain (\ref{conse}). The $Q_N$-conserving functions must be built of Hermitian forms of type (\ref{kk1}) so that they are invariant under
$\psi\to\psi'$. We could certainly repeat the proof perturbatively, as we did to prove shape independence
 in Sec. \ref{shape}
(some issues are resolved by renormalization \cite{peskin}).

\subsection{Hubbard-Stratonovich transformation and renormalization}
\label{rena}

Here we discuss the two techniques often used in context of path integrals. The Hubbard-Stratonovich transformation \cite{hubst}
allows to reduce order of terms in path integral by an auxiliary bosonic field $X(t)$,
\be
\rme^{\int \rmi\rmd t Y^2(t)/4a}\propto\int DX \rme^{\int \rmi \rmd t (\sqrt{a}X(t)Y(t)-X^2(t))}.
\ee
Note that $\sqrt{a}$ is \emph{imaginary} if $a<0$ and in this case $X$ does not correspond to any physical field although
path integrals can be still performed. The renormalization is used to kill divergences appearing in quantum field theory \cite{qft}.
It requires introducing unphysical fields, too. The most common form of renormalizing terms in Largangian is
\be
L_R=f(X_k,\rmi\tilde{X}_m)
\ee
where $\tilde{X}_m$ is the unphysical field as it appears with $i$. Both techniques are fully consistent with
CTP formalism.  For the purpose of this paper - the proof of Lorentz invariance, it is more convenient to use Pauli-Villars
renormalization scheme, with up to several renormalizing fields with large masses (e.g. two bosonic and two fermionis in the case of vacuum polarization) \cite{bjodr}, because dimensional regularization makes Lorentz transformations undefined.
For details of these techniques we refer readers to textbooks \cite{qft,peskin,itzyk}.
\section{Lorentz transformations} 
\label{lorb}
\renewcommand{\theequation}{C.\arabic{equation}}
\setcounter{equation}{0}
\renewcommand{\thesubsection}{(\alph{subsection})}
\setcounter{subsection}{0}
For general multi-indexed relativistic structures we adopt Einstein convention $A\cdot B=A^\mu B_\mu=\sum_{\mu=0,1,2,3}A^\mu B_\mu$ and $A^2=A\cdot A$. 
We denote Minkowski metric tensor
$g^{\mu\nu}=g_{\mu\nu}$ equal $+1$ for $\mu=\nu=0$, $-1$ for $\mu=\nu=1,2,3$ and $0$ otherwise. Next, $g$ is used to lower/raise an index,
$A_\mu=g_{\mu\nu}A^\nu$, $A^\mu=g^{\mu\nu}A_\nu$. In this notation $g^\mu_\nu=\delta^\mu_\nu$ is equal $+1$ if $\mu=\nu$ and $0$ otherwise
and $A^0=A_0$ while $A^{1,2,3}=-A_{1,2,3}$. 
The four-vector measure is denoted $\rmd^4x=\rmd x^0\rmd x^1\rmd x^2\rmd x^3$. We extract the usual three-vectors $\mathbf x=(x^1,x^2,x^3)$, its square
$\mathbf x^2=(x^1)^2+(x^2)^2+(x^3)^3$ and measure $\rmd^3x=\rmd x^1\rmd x^2\rmd x^3$.
The natural Lorentz transformation is defined on four-vectors,
\be
A^{\prime\mu}={\Lambda^{\prime\mu}}_\nu A^\nu\label{lor1}.
\ee
The transformation must satisfy conservation of metric tensor
\be
g^{\mu\nu}={\Lambda^{\prime\mu}}_\alpha{\Lambda^{\prime\nu}}_\beta g^{\alpha\beta}.
\ee
In general, the continuous Lorentz transformation can be written in the form
\be
\Lambda=\exp(-\rmi\omega_{\mu\nu}J^{\mu\nu}/2)
\ee
where $J$ is the generator of the Lorentz group, satisfying
\be
[J^{\mu\nu},J^{\alpha\beta}]=\rmi(g^{\nu\alpha}J^{\mu\beta}-g^{\mu\alpha}J^{\nu\beta}-g^{\nu\beta}J^{\mu\alpha}+g^{\mu\beta}J^{\nu\alpha})
\ee
which yields $J^{\mu\nu}=-J^{\nu\mu}$.
The generator for (\ref{lor1}) reads
\be
{(J^{\prime\mu\nu})^\alpha}_\beta=\rmi(g^{\mu\alpha}g^\nu_\beta-g^{\nu\alpha}g^\mu_\alpha)
\ee
For the scalar field
\be
\Lambda_0 A(x)=A(\Lambda^{\prime -1}x)
\ee
with the corresponding generator
\be
J^{\mu\nu}_0=\rmi(x^\mu\partial^\nu-x^\nu\partial^\mu)\label{j00}
\ee
If the field has vector structure, e.g. $A_\mu(x)$ then the generator reads $J'+J_0$, acting both on argument $x$ and 
four-vector (indices $\mu$). The scalar fields are examples of the general family of Lorentz scalars, combinations of fields that transform according
to $\Lambda_0$. Other examples are \newline $A^\mu(x)B_\mu(x)$, $A(x)\partial_\mu B^\mu(x)$ and $\partial^\mu A\partial_\mu B$
and their combinations (products, sums, multi-indexed forms). In general, if all spacetime indices ($\mu$) appear in pairs, then we have a Lorentz scalar.

Apart from continuous transformations there exist two special discrete transformations, time reversal $x^0\to-x^0$ and parity (mirror) inversion
$x^k\to-x^k$, $k=1,2,3$. One can also make charge conjugation $e\to-e$ (the interaction strength). 
The proof of Lorentz invariance of zero temperature vacuum will be valid
only for continuous transformations. If Lagrangian is invariant with respect to parity inversion or charge conjugation then we can include it, too (which is e.g. 
not the case for weak interactions).

Fermionic fields are represented by four-component \newline spinors $\psi_a$, $a=1,2,3,4$, which have nothing to do with four-vectors.
To define corresponding Lorentz transformations, we first denote Dirac $4\times 4$ matrices, $\gamma^\mu$, with the property
\be
\{\gamma^\mu,\gamma^\nu\}=2g^{\mu\nu}
\ee
and $\gamma^0=\gamma^{0\dag}$, $\gamma^k=-\gamma^{k\dag}$, $k=1,2,3$.
The Lorentz generator reads,
\be
J^{\mu\nu}_s=\rmi[\gamma^\mu,\gamma^\nu]/4.
\ee
The Dirac matrices transform as four-vectors,
\be
\Lambda_s^{-1}\gamma^\mu\Lambda_s=\Lambda^{\prime\mu}_\nu\gamma^\nu.
\ee
The conjugate spinor, $\psi^\dag$ transforms as
$\psi^\dag\Lambda^\dag_s$. Unfortunately, $J_s$ is not Hermitian and $\Lambda^\dag_s\Lambda_s\neq 1$ but
$\Lambda^\dag_s\gamma^0\Lambda_s=\gamma_0$ or $J^{\mu\nu\dag}_s\gamma^0=\gamma^0 J^{\mu\nu}_s$. Therefore we denote
$\bar\psi=\psi^\dag\gamma^0$ and $\bar{\psi}\phi$ is again a Lorentz scalar as well as $A_\mu\bar{\psi}\gamma^\mu\psi$,
$\bar{\psi}\gamma^\mu\partial_\mu\phi$ and so on. General rules to construct a Lorentz scalar are:
\begin{itemize}
\item a product/sum of Lorentz scalars is a Lorentz scalar,
\item $\bar\psi\phi$ and $A$ are Lorentz scalars,
\item $\bar\psi\gamma^\nu\phi$ and $A^\nu$ and $\partial^\nu A$ are Lorentz vectors,
\item $\bar\psi\gamma^\nu\gamma^\mu\phi$ and $\partial^\mu A^\nu$ are Lorentz tensors, etc.,
\item A product of Lorentz vectors (tensors) is a Lorentz scalar if the Greek indices always come in raised/lowered pairs ${}^\mu_\mu$
and Einstein summation is performed.
\end{itemize}

Instead of the Lagrangian, in relativistic spacetime we will rather use its density $\mathcal L$ with the property
\be
L=\int \rmd^3x\;\mathcal L,
\ee
where $\rmd^3x=\rmd x^1\rmd x^2\rmd x^3$ is the three dimensional volume measure.
It allows to write action in the spacetime
\be
\int \rmd t\; L=\int \rmd^4x\,\mathcal L.
\ee
Every Lorentz transformation $\Lambda$ conserves measure \newline $\Lambda(\rmd^4x)=\rmd^4x$.

In the Lagrangian density (and other functionals) one can find every kind of combination of fields, $A(x)$, $B^\mu(x)$, $\psi$, etc.
The Lagrangian will be Lorentz-invariant if its density transforms as Lorentz scalar
\be
\mathcal L(\Lambda A,\Lambda B,\Lambda\psi)=\Lambda_0 \mathcal L(A,B,\psi)
\ee
the generic form of such Lagrangian reads
\ba
&&\sum_{kl}(a^0_{kl}\bar \psi_ki\gamma^\mu\partial_\mu\psi_l+a^1_{kl}(\partial_\mu A_k)(\partial^\mu A_l)+\nonumber\\
&&a^2_{kl}(\partial_\mu B^\mu_k)(\partial_\nu B^\nu_l)+a^3_{kl}(\partial_\mu B^\nu_k)(\partial^\mu B_{\nu l})+
\\
&&+a^4_{kl}A_kA_l+a^5_{kl}B^\mu_kB_{\mu l}+a^6_{kl}\bar{\psi}_k\psi_l+a^7_{kl}B_k^\mu\partial_\mu A_l)+\nonumber\\
&&\sum_{klm}(a^8_{klm}A_m\bar{\psi}_l\psi_m+a^9_{klm}B_{\mu k}\bar{\psi}_l\gamma^\mu\psi_m\nonumber\\
&&+a^{10}_{klm}B^\mu_kB^\nu_l\partial_\mu B_{\nu m}
+a^{11}_{klm}\partial_\mu A_k\bar{\psi}_l\gamma^\mu\psi_m)+\nonumber\\
&&\sum_{klmn}(a^{12}_{klmn}\bar{\psi}_k\psi_l\bar{\psi}_m\psi_n+a^{13}_{klmn}B_k^\mu B_{\mu l}B_k^\nu B_{l\nu})\nonumber
\ea
and the list may be incomplete.

When considering weak interactions, one should include also pseudoscalars $\bar{\psi_k}\gamma^5\psi_l$, $A_\mu\bar\psi_k\gamma^5\gamma^\mu\psi_l$
where $\gamma^5=i\gamma^0\gamma^1\gamma^2\gamma^3$, which transforms as scalar under continuous transformations but changes sign
under parity inversion.
%

\begin{thebibliography}{}
\bibitem{cpt}
R. F. Streater and A. S. Wightman, \textit{PCT, spin and statistics, and all that} (Benjamin, New York, 1964)
\bibitem{jost}
R. Jost \textit{The general theory of quantized fields}  (American
Mathematical Society, Providence, 1965)
\bibitem{cole}
S. Coleman, J. Math. Phys. \textbf{7} (1966) 787
\bibitem{segal} 
I. Segal, in: \textit{Differential Geometry, Group Representations, and Quantization
Lecture Notes in Physics Volume 379} eds. J. Hennig, W. L\"ucke, J. Tolar (Springer, Berlin, 1991) 137-143
\bibitem{koks}
J.F. Koksma and T. Prokopec, arXiv:1105.6296
\bibitem{lors}
J. Schwinger, Phys. Rev. \textbf{82} (1951) 914.
\bibitem{qft}
N. Bogoliubov, A.A. Logunov, A.I. Oksak, I.T. Todorov, \textit{General Principles of Quantum Field Theory}, (Kluwer Academic Publishers, Dordrecht, 1990); L.H. Ryder, \textit{Quantum Field Theory} (Cambridge University Press, 1985).
\bibitem{weinb}
S. Weinberg, \textit{The Quantum Theory of Fields}, (Cambridge University Press, Cambridge, 1995).
\bibitem{peskin}
M. Peskin, D. Schroeder, \emph{An Introduction to Quantum Field Theory}, (Perseus Books, Reading, 1995).
\bibitem{bell-book}
M. Le Bellac, \textit{Thermal Field Theory} (Cambridge University Press, Cambridge, 2000).
\bibitem{nair}
V.P. Nair, G.D. Sprouse, private communication (2013). 
\bibitem{kost}
V. A. Kostelecky,
Phys. Rev. D \textbf{69} (2004) 105009. 
\bibitem{schwinger}
J. Schwinger, J. Math. Phys. {\bf 2} (1961) 407.
\bibitem{kaba}
L.P. Kadanoff and G. Baym, \textit{Quantum Statistical Mechanics}, (W.A. Benjamin, New York, 1962).
\bibitem{kel}
L. Keldysh, Sov. Phys. JETP \textbf{20} (1965) 1018.
\bibitem{our}
A. Bednorz, K. Franke and W. Belzig, New. J. Phys. \textbf{15} (2013) 023043.
\bibitem{feyn}
R.P. Feynman and A.R. Hibbs, \textit{Quantum Mechanics and Path Integrals} (McGraw-Hill, New York,1965)
\bibitem{mats}
T. Matsubara, Prog. Theor. Phys. \textbf{14} (1955) 351.
\bibitem{kap}
 J. I. Kapusta and C. Gale, \textit{Finite-Temperature Field Theory}, 
(Cambridge University Press, Cambridge, 2005).

\bibitem{bam}
K. T. Mahanthappa, Phys. Rev. \textbf{126} (1962) 329; P. M. Bakshi and K. T. Mahanthappa, J. Math. Phys. \textbf{4} (1963) 1, 12 ; R. A. Craig, J. Math. Phys. \textbf{9} (1968) 605.
\bibitem{mills}
R. Mills, \textit{Propagators for many-particle systems}, (Gordon and Breach, New York, 1969).
\bibitem{calz}
E. Calzetta and B. L. Hu,
Phys. Rev. D \textbf{35} (1987) 495; \textbf{37} (1988) 2878. 
\bibitem{wein}
S. Weinberg, Phys. Rev. D \textbf{72} (2005) 043514.
\bibitem{kobeskow}
R. L. Kobes, K. L. Kowalski, Phys. Rev. D \textbf{34} (1986) 513.
\bibitem{kra}
U. Kraemmer and A. Rebhan,
Rep. Prog. Phys. \textbf{67} (2004) 351.
\bibitem{mabi}
M. Le Bellac and H. Mabilat, Phys. Lett. B,  \textbf{381} (1996) 262;
H. Mabilat, Z. Phys. C  \textbf{75} (1997) 155.
\bibitem{kms}
R. Kubo, J. Phys. Soc. Jpn. \textbf{12}, 570 (1957); Rep. Prog. Phys. \textbf{29} (1966) 255 ;
P. C. Martin and J. Schwinger, Phys. Rev. \textbf{115} (1959) 1432.
\bibitem{ksw}
R. L. Kobes, G. W. Semenoff and N. Weiss, Z. f. Physik C \textbf{29} (1985) 371.
\bibitem{nieg}
A. Niegawa, Phys. Rev. D \textbf{40} (1989)  1199.
\bibitem{evans}
T. S. Evans,
Phys. Rev. D \textbf{47} (1993) R4196.
\bibitem{gelis}
F. Gelis, Phys. Lett. B \textbf{455} (1999) 205;
Z. Phys. C,  \textbf{70} (1996) 321.
\bibitem{itzyk}
C. Itzykson, J.-B. Zuber, \textit{Quantum Field Theory} (McGraw-Hill, New York, 1980).
\bibitem{furnser}
R. J. Furnstahl and B. D. Serot,
Phys. Rev. C \textbf{44} (1991) 2141.
\bibitem{weld}
H.A. Weldon, Phys. Rev. D \textbf{26} (1982) 1394.
\bibitem{hanggi}
J. Dunkel, P. H{\"a}nggi, and S. Hilbert, Nature Phys. \textbf{5} (2009) 741.
\bibitem{niemi}
A.J. Niemi, G.W. Semenoff, Nucl. Phys. B \textbf{220} (1984) 181; Ann. Phys. \textbf{152} (1984) 105.
\bibitem{kobes2}
R. L. Kobes and G. W. Semenoff, Nucl. Phys. B  \textbf{260} (1985) 714; \textbf{272} (1986) 329.
\bibitem{landweert}
N.P. Landsman and C.G. van Weert, Phys. Rep. \textbf{145} (1987) 141.
\bibitem{fdt}
H. B. Callen and T. A. Welton, Phys. Rev. \textbf{83} (1951) 34.

\bibitem{chou}
K. Chou, Z. Su, B. Hao and L. Yu, Phys. Rep. \textbf{118} (1985) 1.
\bibitem{jackiw}
L. Dolan, R. Jackiw, Phys. Rev. D \textbf{9} (1974) 3320.
\bibitem{evan2}
T. S. Evans and A. C. Pearson, Phys. Rev. D \textbf{52} (1995) 4652.
\bibitem{nieg06}
M. Inui, H. Kohyama, and A. Niegawa, Phys. Rev. D \textbf{73} (2006) 047702.
\bibitem{gem}
M. Gell-Mann and F. Low, Phys. Rev. \textbf{84} (1951) 350
\bibitem{kame}
A. Kamenev, A. Levchenko, Adv. in Phys. \textbf{58} (2009) 197.
\bibitem{wick}
G.C. Wick, Phys. Rev. \textbf{80} (1950) 268.
\bibitem{wickt}
T. S. Evans, D. A. Steer, Nucl. Phys. B \textbf{474} (1996) 481.
\bibitem{zinn}
J. Zinn-Justin, \textit{Path Integrals in Quantum Mechanics} (Oxford University Press, Oxford, 2005).
\bibitem{hubst}
R. L. Stratonovich, Sov. Phys. Dokl. \textbf{2} (1958) 416.
J. Hubbard, Phys. Rev. Lett. \textbf{3} (1959) 77.
\bibitem{bjodr}
J.D. Bjorken, S.D. Drell, \textit{Relativistic Quantum Mechanics} (New York: McGraw-Hill, New York, 1998)

\end{thebibliography}
%

\end{document}